\documentclass[runningheads]{llncs}
\usepackage{graphicx}
\usepackage{soul}
\usepackage[utf8]{inputenc} %unicode support
\usepackage{url}
\begin{document}
\title{Prioritization of COVID-19-related literature via unsupervised keyphrase extraction and document representation learning}
\titlerunning{Prioritization of COVID-19-related literature}
\author{Bla\v{z} \v{S}krlj\inst{1,2}\orcidID{0000-0002-9916-8756} \and
Marko Juki\v{c}\inst{3} \and Nika Er\v{z}en \and \\
Senja Pollak\inst{1}\orcidID{0000-0002-4380-0863} \and Nada Lavra\v{c}\inst{1,2}}
\authorrunning{\v{S}krlj et al.}
\institute{Jo\v{z}ef Stefan Institute, Ljubljana, Slovenia \and
Jo\v{z}ef Stefan International Postgraduate School, Ljubljana, Slovenia \and
Faculty of Chemistry and Chemical Technology, University of Maribor, Slovenia}
\maketitle              % typeset the header of the contribution
\begin{abstract}
The COVID-19 pandemic triggered a wave of novel scientific literature that is impossible to inspect and study in a reasonable time frame manually. Current machine learning methods offer to project such body of literature into the vector space, where similar documents are located close to each other, offering an insightful exploration of scientific papers and other knowledge sources associated with COVID-19. However, to start searching, such texts need to be appropriately annotated, which is seldom the case due to the lack of human resources. In our system, the current body of COVID-19-related literature is annotated using unsupervised keyphrase extraction, facilitating the initial queries to the latent space containing the learned document embeddings (low-dimensional representations). The solution is accessible through a web server capable of interactive search, term ranking, and exploration of potentially interesting literature. We demonstrate the usefulness of the approach via case studies from the medicinal chemistry domain.
\end{abstract}
\fbox{Published version at
\url{https://link.springer.com/chapter/10.1007/978-3-030-88942-5_16}}
\section{Introduction}
Severe acute respiratory syndrome coronavirus 2 or SARS-CoV-2 is a coronavirus, member of the \emph{Coronaviridae} family, a positive-sense single-stranded (+ssRNA) RNA virus~\cite{Nat2020}.
%\cite{Nat2020,Zhu2020,Cui2018}. 
The novel virus (initially 2019-\emph{n}CoV now named SARS-CoV-2; \emph{‘n’} - novel) was reported in December of 2019 to be originating from Wuhan, Hubei China~\cite{Wu2020}. In the closing of 2019-early 2020, the virus caused a global pandemic of the COVID-19 disease~\cite{Wang2020}.
The latter is of grave concern, as the majority of cases display mild symptoms, but up to 15 \% of patients progress to pneumonia and multi-organ failure leading to potential death, especially without medical assistance~\cite{Wang_2020_jama}.

While there are no registered drugs, but several drug and vaccine discovery programs are being actively developed and scaled up, the scientific community coherently responded to the COVID-19 pandemic resulting in 
%the amassment of data. Thus a non-trivial research venue was unveiled ~\cite{brainard2020scientists,wynants2020prediction}, characterized by 
an increasing amount of literature that is beyond the search capabilities of individual medical professionals~\cite{wang2020cord}. 
Exploration of scientific literature can be facilitated by computationally feasible approaches to summarizing a large amount of text~\cite{jones2002interactive}.
This work explores how unsupervised document representation learning and keyphrase extraction methodologies ~\cite{hasan2014automatic,campos2020yake} can be used to build a fast, scalable web server suitable for \textbf{literature prioritization}. 
To this end, we implemented a web server tool and showcase the solution's scalability on one of the largest currently known collections of COVID-19-related full medical document databases -- CORD19~\cite{wang2020cord}.

We next present the related work, followed by the developed web server and its use cases. We conclude with a discussion of the developed tool and further work.

\section{Related work}

With the introduction of freely available literature, multiple tools have been recently developed~\cite{Hutson2020}. 
%We summarise the tools next and compare them to the proposed web server. 

%In their recent work~\cite{bras2020visualising}, the authors explore how the body of COVID-19 literature could be visualized via a bubble-like visualization, where the main keywords are grouped.
Bras et al.~\cite{bras2020visualising-2}  propose bubble-like visualization of the COVID-19 literature using keyword groups, resulting in hundreds of documents retrieved. The tool offers search based on pre-defined sets of keywords, which can be ambiguous and potentially result in papers not directly related to a given query, offering a fast overview of key topics. 

Another interesting project is the Watson Annotator of Clinical data\footnote{\url{https://www.ibm.com/cloud/watson-annotator-for-clinical-data}}, capable of highlighting key terms within a given document. This tool aims not to provide the global search across the literature but to annotate an e.g., copy-pasted document with named entities. Such annotation can be very useful for medical professionals, as it offers, similarly to this work, quick insights into the key concepts appearing in a given document. A substantially different approach was undertaken by Google\footnote{\url{https://covid19-research-explorer.appspot.com/}}, where a question answering regime was adopted. Their search engine can identify publications based on a natural language-based query, e.g., ``What is the medical care for patients during COVID-19 epidemic?''. The engine recommends (according to its internal ranking) the documents that are of potential interest. 

An interesting approach is also CADTH COVID-19 pandemic online tool\footnote{\url{https://covid.cadth.ca/literature-searching-tools/cadth-covid-19-search-strings/}}, which offers string search to topics related to COVID-19. Another recently released tool is the COVIDScholar\footnote{\url{https://covidscholar.org/}} whose core functionality is the most similar to the tool presented in this paper. It is based on word and document embedding techniques used for semantic search. It leverages open data from various data sources. The main results are links to full papers with abstracts and the most similar documents via document embeddings. The tool, however, does not explore the possibility of full-text annotation via keyphrase extraction, which is among the key functionalities of our tool. 
Finally, the SPIKE tool by the Allen Institute
also offers an exploration of documents at scale, offering insight into named entities and their relations within documents, which can be very useful when attempting to answer specific queries based on literature\footnote{\url{https://spike.covid-19.apps.allenai.org/datasets/covid19/search}}. The data set that gave rise to the tools was initially offered at Kaggle\footnote{\url{https://www.kaggle.com/allen-institute-for-ai/CORD-19-research-challenge}}. 
The majority of the available data on the heavily studied SARS-CoV-2 topic and related COVID-19 pandemic resides on traditional literature bodies that employ heavy user involvement and literature study. A few examples of the literature bodies besides the aforementioned CORD19 database are offered by NIH as SARS-CoV-2 Resources\footnote{\url{https://ncbi.nlm.nih.gov/sars-cov-2/}}, NCBI as LitCovid\footnote{\url{https://nncbi.nlm.nih.gov/research/coronavirus/}}, Rutgers university as COVID-19 Information Resources\footnote{\url{libguides.rutgers.edu/covid19\_resources/}}, ECDC\footnote{\url{ecdc.europa.eu/en/coronavirus}}, WHO\footnote{\url{search.bvsalud.org/global-literature-on-novel-coronavirus-2019-ncov/}}\footnote{\url{who.int/emergencies/diseases/novel-coronavirus-2019}}, USCF\footnote{\url{guides.ucsf.edu/COVID19/literature}}, Wiley\footnote{\url{novel-coronavirus.onlinelibrary.wiley.com}}, ACS\footnote{\url{acs.org/content/acs/en/covid-19.html}} and others. A more comprehensive overview of related literature databases and tools~\cite{10.1093/bib/bbaa296} is provided by CDC\footnote{\url{cdc.gov/library/researchguides/2019novelcoronavirus}}.

%The tabular summary of the related work is given as Table~\ref{tab:summary-related}.

%x\hl{dobro bi bilo dodati kako delo za primerjavo in kaj Covid-19 explorer dodatnega omogoca, ce je seveda kaj, kar gre nam v plus., v komentarju nekaj linkov}
%https://arxiv.org/pdf/2005.06380.pdf
%https://acd-try-it-out.mybluemix.net/preview %https://covid19.ccg.unam.mx/info-lcovid.html,https://covid19-research-explorer.appspot.com/ %https://www.nature.com/articles/d41586-020-01733-7}}

\section{COVID-19 Explorer design, implementation and functionality}
\label{sec:implementation}

%The overview of the proposed web server architecture is shown in Figure~\ref{fig:architecture} and is discussed next.

\begin{figure}[h!]
    \centering
        \vspace*{-1.5cm}
    \includegraphics[width = 1\linewidth]{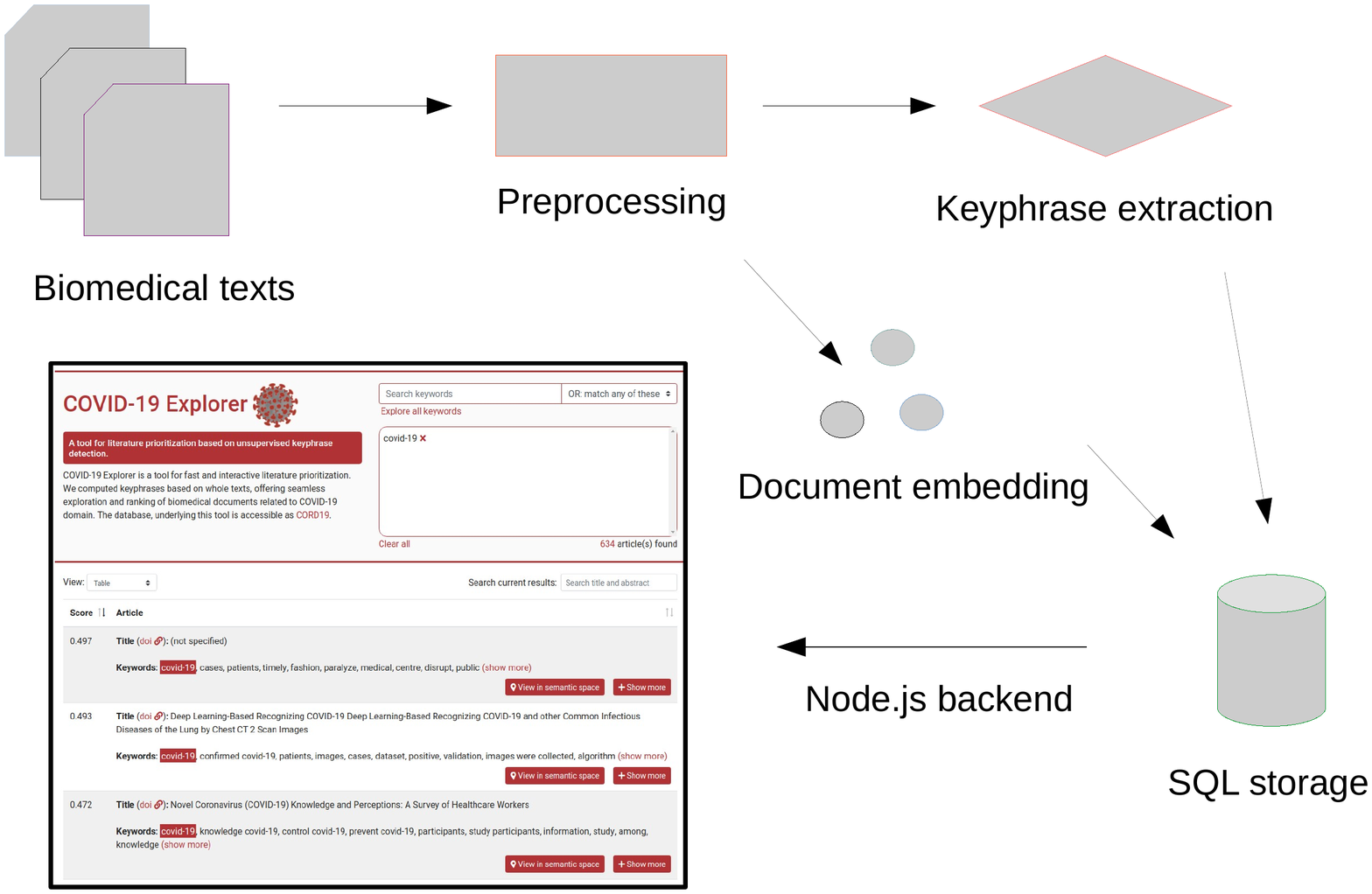}
    \vspace*{-1.5cm}
    \caption{Visualization of the main processing steps considered by the proposed solution.}
    \label{fig:architecture}
\end{figure}

The proposed COVID-19 Explorer webserver architecture,  shown in Figure~\ref{fig:architecture}, is comprised of two main parts.

First, the raw body of COVID-19-related literature is preprocessed and stored in the form, suitable for the two subsequent machine learning tasks. The first task, keyphrase extraction, is conducted with the recently introduced RaKUn algorithm~\cite{rakun}, additionally equipped with scientific stop-word lists to prevent noisy keyphrases from being detected. The second task, document representation learning, is conducted by using the widely adopted doc2vec document embedding algorithm~\cite{10.5555/3044805.3045025}, used to learn representations of abstracts of individual documents. Once keyphrases and document embeddings are obtained, they are stored in a form suitable for fast access. The document embeddings are also projected to 2D with UMAP~\cite{McInnes2018}, a non-linear dimensionality reduction tool, as the front end part of the webserver offers \textbf{interactive exploration} also by querying the semantic (2D) space directly.

All the information is presented in the form of a responsive and fast front end, requiring minimal computational resources on the client-side.

\begin{figure}[h!]
    \centering
     \includegraphics[width = \linewidth]{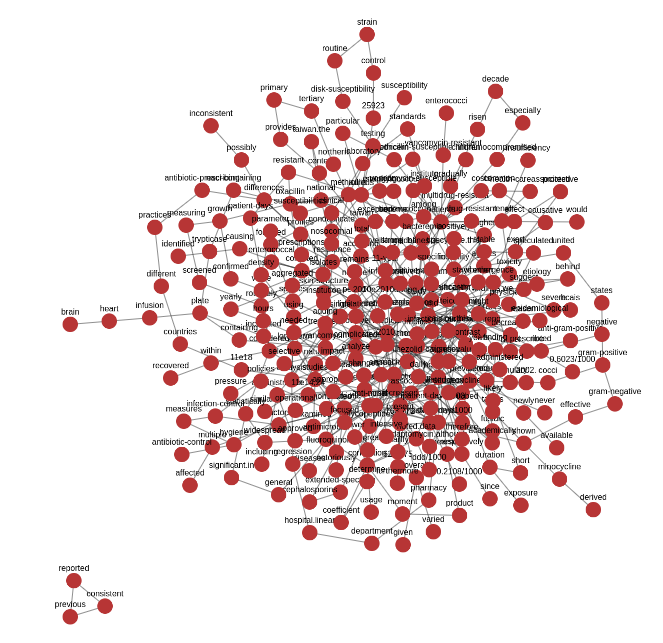}
                 \vspace*{-0.5cm}
    \caption{An example RaKUn token graph. Each node represents a token in the document. Documents, when linked together, form the whole document graph suitable for identifying the keyphrases (as paths in this graph).}
    \label{fig:tokens}
                     \vspace*{-0.1cm}
\end{figure}

\subsection{Keyphrase extraction}
\label{subsec:keyphrases}

One of the key functionalities of the COVID-19 Explorer is that it enables a direct search via keyphrases, computed from \emph{whole} scientific documents (papers, reports, etc.). The extraction method is the in-house developed RaKUn algorithm ~\cite{rakun}. The algorithm first transforms a given collection of sentences (a document) into a \emph{document graph} -- a graph comprised of key tokens, linked via the co-occurrence relation. An example graph is shown in Figure~\ref{fig:tokens}.
% \hl{naredi en figure ko je zoom na ene 3-4 node pa se da vse 3-4 interpretirat - varianta - vesolje - puščica zoom na 3 node in spodaj opis vseh 3}

\begin{figure}[t!]
    \centering
    \includegraphics[width=\linewidth]{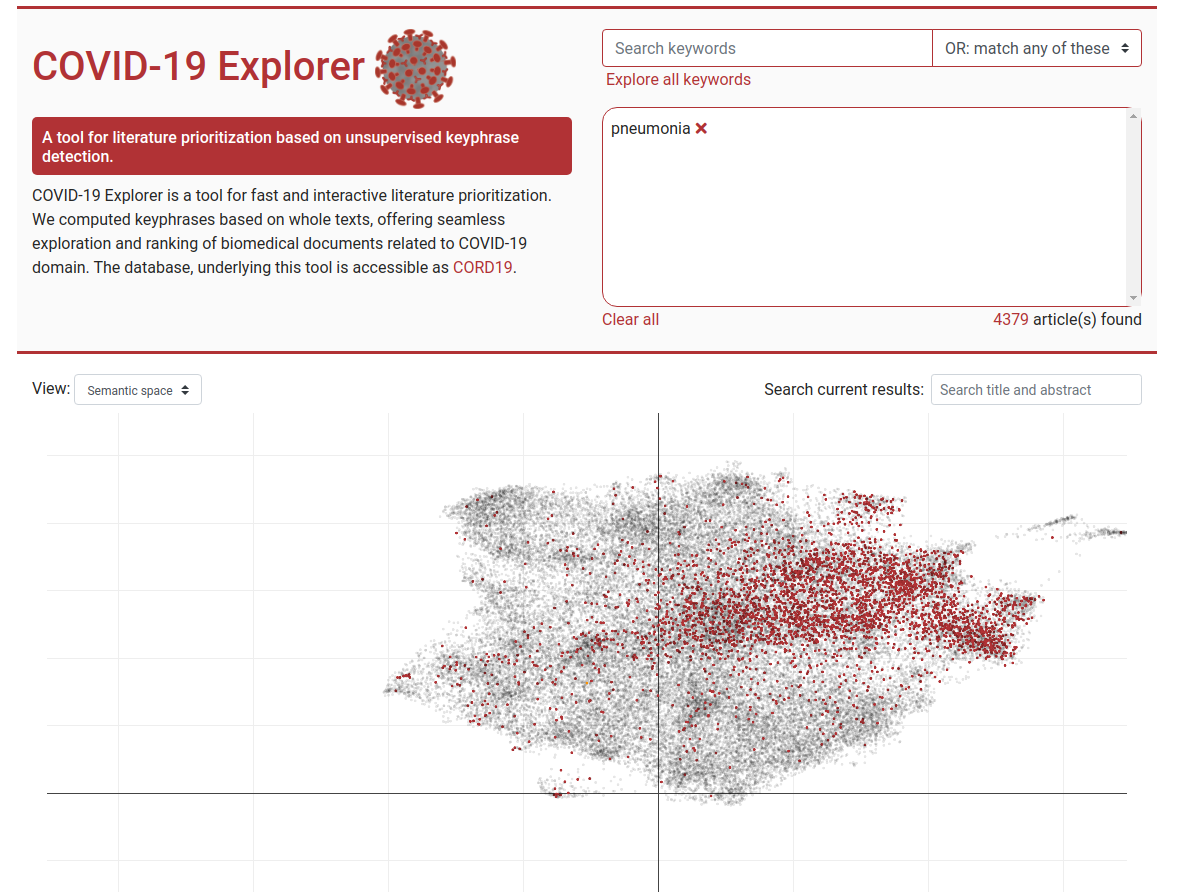}
    \caption{Global interactive space of document embeddings.}
    \label{fig:keyphrase-prioritization}
\end{figure}

Once the graph is constructed for a given collection of text, \emph{ranking} of nodes is performed to identify single, two, and three-term \emph{keyphrases}. The webserver also implements an auto-suggestion option, which offers interactive exploration of possible search queries in real-time.
%\hl{Lahko bi tudi poudaril, da dela tudi auto complete in za dan zacetek searcha predlaga keyworde iz celega korpusa. Men se zdi to koristna funkcionalnost - poglej sliko KeywordProposed (ni v  main vkljucena, samo v projektu)}
The current implementation of RaKUn employs \emph{load centrality}, a centrality measure based on the amount of shortest paths that pass through a given node. The keyphrase computation step is conducted in parallel for each of the considered documents. The resulting keyphrases and the underlying token graphs are stored and browsed interactively as a part of the front-end functionality.
Further, the keyphrase extraction %%%\textbf{keyphrase extraction} 
offers another functionality that is crucial in the considered document prioritization task -- each keyphrase has a dedicated score for a given document, meaning that the documents themselves can be prioritized for the global keyphrase score. An example of how this space can be directly inspected is shown in Figure~\ref{fig:keyphrase-prioritization}. 
In addition to scoring a given keyphrase within a given document, the search results also show other keyphrases and the document title linked to the corresponding DOI.

\begin{figure}[h!]
    \centering
    \includegraphics[width=\linewidth]{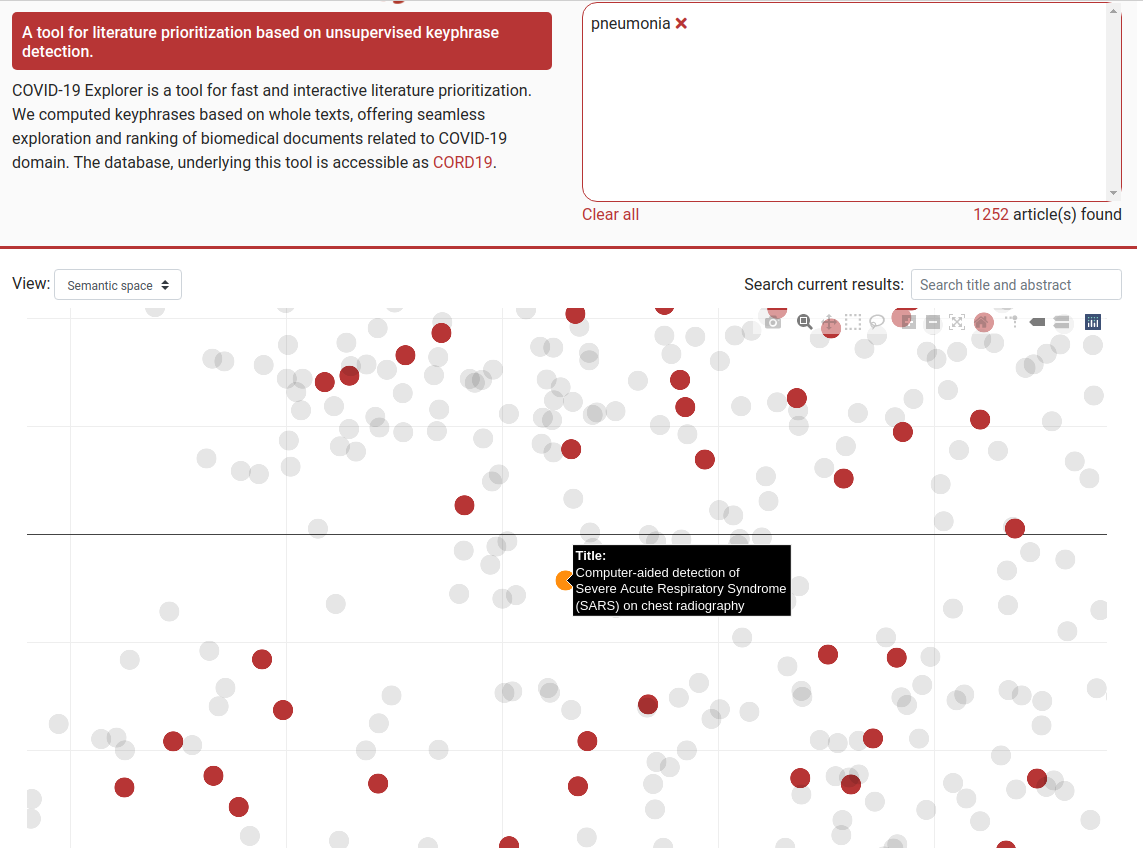}
    \caption{The COVID-19 Explorer's semantic viewer. Each point in the shown space represents a document. The positions in the \emph{global} space of documents are determined based on the distances between the representations of the documents' abstracts. Intuitively, the documents close to one another can offer insight into the semantically similar document. The implemented viewer offers a direct exploration of documents -- each point is clickable and triggers an element with a detailed description of a given document.}
    \label{fig:viewer}
\end{figure}

\subsection{Global document representation learning}
\label{sec:representation-learning}
A prominent capability of the natural language processing methods developed in recent years is that of \emph{learning} the representation of a collection of texts, instead of merely considering the set of hand-crafted features. The current implementation of the COVID-19 Explorer exploits the widely used doc2vec algorithm~\cite{10.5555/3044805.3045025} to learn the representations of \emph{every} document abstract. The purpose of this step is to map the considered collection of documents into the same semantic %%%\textbf{semantic} 
space, offering the capability to explore the e.g., semantic neighborhoods of a given document, \emph{interactively}.
The current implementation of the COVID-19 Explorer first computes 256-dimensional representations of individual abstracts and next projects them to two dimensions via the UMAP~\cite{McInnes2018} tool that approximates a low dimensional manifold representative of the learned high dimensional space. The implemented \emph{semantic viewer} is shown in Figure~\ref{fig:viewer}.

\section{Case studies}
\label{sec:results}
\begin{figure}
    \centering
    \includegraphics[width=\linewidth]{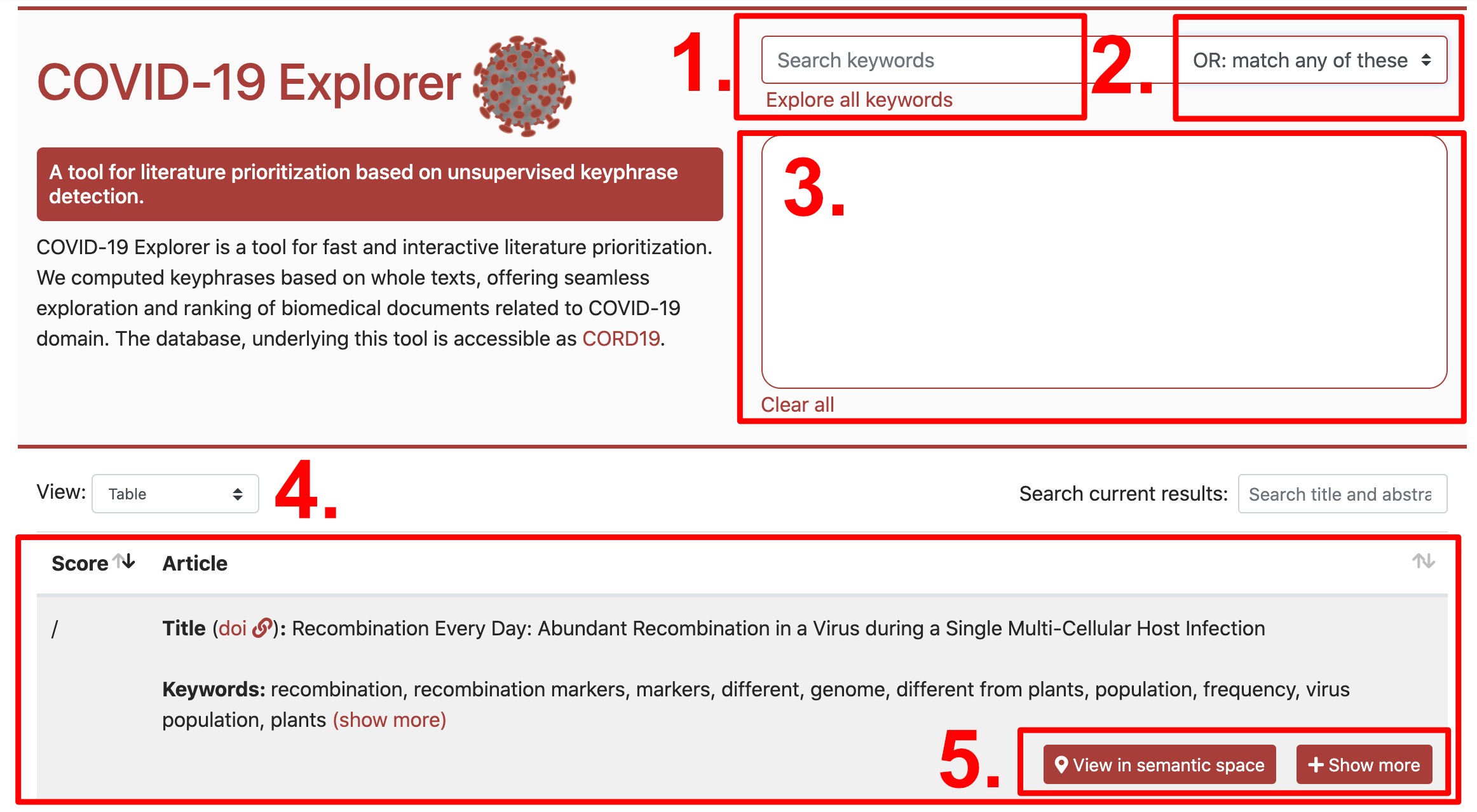}
    \caption{The COVID-19 Explorer's welcome screen and user input fields. Relevant key sections of the online tool are emphasized and numbered in red color: 1. user keyword inspection and input field, 2. Boolean operators imposed on keywords, 3. selected keyword inspection window, 4. Dynamically updating result field displaying semantically related peer-review articles, 5. specific article detail button and 2D visualization of the corresponding semantic space. }
    \label{fig:uhelp}
\end{figure}

This section presents the application of the reported webserver tool and its main functionalities. We also showcase its performance on multiple use-cases, aimed at the fast and efficient hypothesis elaboration for research on COVID-19 drug design. We present the general research-field examination together with key scientific questions regarding the development of novel drugs against the SARS-CoV-2 pathogen and SARS-CoV-2 therapeutic target examination. In the presented cases, we demonstrate how the COVID-19 Explorer effectively identifies the relevant semantically associated literature. Upon navigating to \textbf{http://covid19explorer.ijs.si/}, the user is presented with a welcome screen where keyword(s) can be chosen (Figure~\ref{fig:uhelp}, subfigure 1) and their relationship using Boolean operators (Figure~\ref{fig:uhelp}, subfigure 2). The user is then presented with a list of examined keywords (Figure~\ref{fig:uhelp}, subfigure 3) and the list of semantically connected articles is dynamically updated in the output field below (Figure~\ref{fig:uhelp}, subfigure 4). Individual pinpointed articles can be examined in detail and its semantic space visualized (Figure~\ref{fig:uhelp}, subfigure 5).

\subsection{Case study 1: General COVID-19 domain inquiry}
\smallskip
Single keyword examination using the term \textbf{pandemic} yields results where the top-scoring article (Score: 0.542) entitled \emph{“The pandemic present”} immediately affords social anthropology discourse on current pandemic threats including COVID-19~\cite{Whitacre2020}, Amongst the 10 top scoring peer-review articles offered by the COVID-19 Explorer, 7 investigate the COVID-19 emergency and 3 articles offer information on the influenza pandemics. The former and latter are two major subjects found in modern medicinal literature regarding the general topic of pandemics~\cite{Gates2020,Kilbourne2006,Chew2010}. The journals found by the COVID-19 Explorer are all of high-impact in the respective fields and encompass \emph{Social Anthropology}, %~\cite{Whitacre2020}, 
\emph{Journal of Medical Humanities}, %~\cite{Bracken2020}, 
\emph{Emerging Infectious Diseases}, %~\cite{Ujike2011}, 
\emph{Public Health}, %~\cite{McConnell2010}, 
\emph{The Canadian Journal of Addiction}, %~\cite{elGuebaly2020}, 
\emph{The Lancet} %~\cite{Shortridge2003,McConnell2005,Boyce2020} 
and \emph{British Journal of Surgery}. %~\cite{Farid2020}. 
The keyword \textbf{pandemic} thus offered a balanced perspective from social studies (20 \%) and modern medical science perspective (80 \%) on the field of pandemic studies, especially focusing on COVID-19 on the first, and influenza in the second place. A similar broad, yet subject-focused outlook in correlation to the present-time problem would be difficult to impossible for identification using one search operation in other peer-review literature search engines.
Supplementing the search with the \textbf{covid} keyword and using AND Boolean operator offered by COVID-19 Explorer shifts the result focus entirely to COVID-19 related peer-review articles. Top scoring hits offer the general outlook on COVID-19 health effects \cite{Randolph2020}, possible treatments \cite{Honore2020}, promoting the mental healthcare during the COVID-19 pandemic   \cite{Novins2020,Advani2020}, elaborating on children study problems  \cite{Agarwal2020} and clinical problems encountered during the COVID-19 pandemic  \cite{Cattaneo2020,Hing2020}. The results in effect mirror the key media-reported problems and challenges imposed by the current global crisis and could be retrieved by COVID-19 Explorer in a single search operation.
\smallskip

\subsection{Case study 2: SARS-CoV-2 potential therapeutic drug/target identification}
\smallskip
There are only a few therapeutic options for SARS-CoV-2, a pathogen causing worldwide havoc~\cite{Jin2020,Tiwari2020}. Therefore, novel drug design is paramount, as well as an inquiry into viral biochemistry along with the identification and assessment of potential novel therapeutic targets that could be of use for the development of novel drugs ~\cite{Wu2020}. Using the COVID-19 Explorer, we examined the subject by using two straightforward keywords, i.e. \textbf{sars-cov-2} and \textbf{receptor} joined by AND operator. The reported web server tool immediately delivered a focused overview on the subject comprised of 10 top scoring articles where peer-review literature offered an outlook on the antimicrobial chemical matter with activity against SARS-CoV-2 virus - a repurposing study in one article \cite{Ijaz2020}, human-to-human transmission of COVID-19 in one article \cite{Mohseni2020}, elaboration on human ACE2 receptor in 7 articles \cite{Lutchman2020,Zhang2020,Panciani2020,Ortega2020}, current insight into viral morphology, biochemistry, and pathogenesis \cite{Kumar2020} and involvement of the virus in resulting COVID-19 disease \cite{Saxena2020}. Worth mentioning is also the correctly identified connection between ACE2 host entry receptor and viral binding partner S-protein \cite{Zhang2020} (COVID-19 Explorer 4th hit article with the score of 0.295). This key finding represents a prominent therapeutic target for the development of novel drugs and vaccines against SARS-CoV-2 ~\cite{Zhou2020}. Modifying the search to \textbf{sars-cov-2} and target % \textbf{target} 
keywords, associates the discourse tightly to medicinal chemistry and results in peer-review article focus on ACE2 in viral pathogenesis \cite{Su2020}, comparison of the SARS-CoV-2 with SARS-CoV and MERS-CoV \cite{Fani2020} as well as identifying two key potential therapeutic targets for the development of novel drugs against SARS-CoV-2 - 3CLpro and RdRp \cite{Buonaguro2020,Li2020}. Furthermore, the top 10 suggested articles of the COVID-19 Explorer tool also include elaboration on the viral entry mechanism involving TMPRSS2 extracellular protease (5th hit with a score of 0.285) \cite{Saxena2020}. TMPRSS2 protease is essential in understanding spike protein processing and the mechanism of viral cell entry \cite{Matsuyama2020}.
\smallskip

\begin{figure}[b!]
    \centering
    \vspace*{-1cm}
    \includegraphics[width=\linewidth]{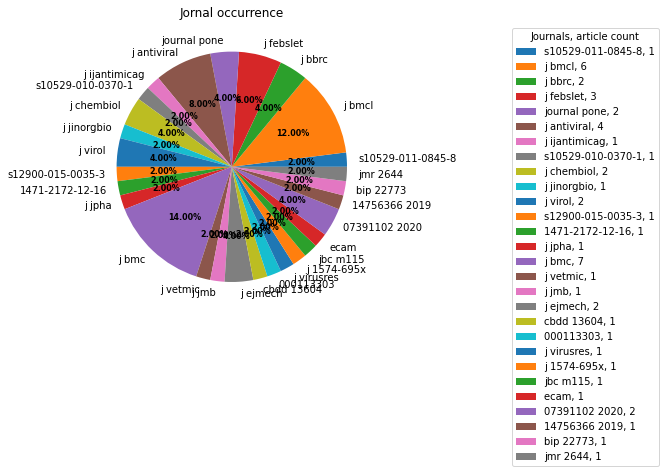}
    \caption{Left: API hitlist for the keyword 3clpro; Right: Article counts from specific journals}
    \label{fig:pie1}
        \vspace*{-1cm}
\end{figure}
\subsection{Using the COVID-19 Explorer API}
\smallskip
For the user benefit, the COVID-19 Explorer web server tool also exposes a RESTful API with a simple syntax:
\begin{center}
\textrm{http://cord19explorer.ijs.si/gp/api?keyword=}\textbf{query} 
\end{center}
where \textbf{query} is a user-defined search term. Additional keywords can be added with \emph{\&query2\&query3…} and the search output hitlist can be limited with a \emph{limit=N} and \textbf{N} is the requested number of articles. To restraint the server load, the current hitlist is limited to 50 but this can be adjusted as needed. For example, using a scripting language (e.g., Python) and perhaps a notebook software (e.g., Jupyter) elaborate search patterns can be performed and results analyzed. For example upon searching for a simple \textbf{keyword}=\textbf{3clpro} a hitlist is obtained and can be readily analysed article/publisher-wise Figure~\ref{fig:pie1}.

Similarly, the user can easily obtain articles with a specific term in the abstract. For example, a list of articles with a term inhibitor in the abstract as a subset of API hitlist, discern the publication year, field of study and so on. The exposed RESTful API is thus an easy approach towards automatization and incorporation of  COVID-19 Explorer's searches into other tools and workflows.

\smallskip

%Two other case studies are presented in the Appendix.

\section{Discussion}
\label{sec:discussion}
Let us discuss the usefulness of the proposed tool alongside its drawbacks. The developed COVID-19 Explorer offers scalable and highly efficient \emph{summarization} of scientific documents via keyphrases, based on \emph{whole} texts. Compared to e.g., conventional approaches are undertaken by large databases such as e.g., PubMed, where the keywords are determined by the authors themselves (and manually tagged by professionals), the purpose of this work was to demonstrate that at least to some extent, this process can be automated in and \emph{unsupervised manner}, without any human interventions, and offers a scalable approach to the exploration of vast amounts of scientific literature. One of the key goals of the COVID-19 Explorer is to \emph{filter} existing information and thus simplify arduous exploration (often random) of scientific literature to domain experts.

The proposed implementation offers two fundamentally different approaches to the exploration of the document, which we were able to link -- namely, the tool offers exploration directly in the space of latent embeddings of documents via an interactive 2D visualization, but also exploration directly via \emph{ranked} keywords, present throughout the documents. Exploration via keywords was specifically optimized by taking into account the existing lists of scientific stopwords. Thus the RaKUn keyphrase extraction phase was adopted for the scientific domain, which is also a contribution of this work. Furthermore, RaKUn was updated with the capability of detecting connectives -- words that link multiple terms; for example ``COVID-19 \textbf{in} the USA'' represents a keyphrase where \textbf{in} was identified to fit between the two key phrases. The RaKUn achieves such behavior by backtracking back to the raw text and statistically identifying how suitable a given connective is. Therefore a suitable context can be tailored upon the search request. Finally, we believe that the existing semantic space could be improved by incorporating both the document embedding information, additionally equipped with the whole-document keywords. Finally, we plan to explore more sophisticated summarisation techniques, summarised for the interested reader in ~\cite{el2021automatic}; many of these techniques are based on computationally more expensive neural language models, which, however, could offer superior performance.

Using the reported tool for general COVID-19 subject outlook revealed a hit-list of semantically connected articles where a broad overview on the global COVID-19 crisis from the social studies and a modern medicinal perspective was obtained in a single search query. Furthermore, simple keyword exploration of the general subjects from the Medicinal chemistry domain quickly afforded relevant high-impacting peer-review articles from journals respective on the field. In essence, a review article deconstruction was achieved with key articles, ideas, and themes offered as top hits. Inspection of specific terms from SARS-CoV-2 antiviral drug design returned lists of relevant primary literature as well as associations to other complementary research approaches, e.g., exploration of 3CLpro target exposed associations to PLpro, RdRp, and 2'-O-MTase therapeutic targets. Furthermore, the reported tool helps the user with an implemented API in the background and precomputed lists of related keywords in the foreground. Upon inputting a specific keyword, a list of %%%\textbf{semantically related suggestions}
semantically related suggestions is offered, unaffected by the user preference but rather derived from the underlying body of data.

\section{Conclusions}
\label{sec:conclusions}
In this work, we presented an approach for summarization of %%%\textbf{large collections of scientific documents based on automatic keyphrase extraction}.
large collections of scientific documents based on automatic keyphrase extraction. The approach was extended with a simple-to-use web interface, where users can explore the semantic space of COVID-19-related medical literature.
As keyphrases were computed based on whole texts automatically, the proposed tool offers exploration capabilities beyond a few author-assigned keywords present in dominant search engines. Furthermore, the keyphrase extraction algorithm was specifically adapted for the biomedical domain via scientific stop word lists, which substantially improved the search performance and the quality of the results.

We demonstrated the usefulness of the proposed approach in different case studies, studying different aspects of the current COVID-19 pandemic, from molecular (receptor) level to more general, disease co-occurrence level. We demonstrated that the tool indeed offers a fast and intuitive exploration of the scientific literature as well as an alternative view on the underlying body of work. Furthermore, the proposed article ranking system, which assigns a score to each paper, was shown to prioritize the literature in a manner suitable for literature-based discovery and exploration. The article ranking idea is also a novelty of this paper.
Finally, even though the existing web service offers an intuitive and fast exploration of existing literature, we believe the approach could be extended to incorporate contextual embeddings, which could further distill the relevant literature. Even though the focus of this work was the CORD-19 corpus, the authors are aware, the proposed approach can be generalized for \emph{any} collection of relevant literature.

\section{Availability and Requirements}
\label{sec:availability}

COVID-19 Explorer is available at http://covid19explorer.ijs.si/ as a freely accessible webserver. The web server's landing page includes the links to the repository and the data used to completely reproduce the webserver locally.

\section*{Acknowledgements}
 This work was supported by the Slovenian Research Agency (ARRS) core research program P2-0103 and the CRP project V3-2033. The work of the first author was financed by the ARRS young researchers grant. The work was also supported by European Union's Horizon 2020 research and  innovation programme under grant agreement No 825153, project EMBEDDIA (Cross-Lingual Embeddings for
Less-Represented Languages in European News Media).
\bibliographystyle{splncs04}
\bibliography{mybibliography.bib}
\end{document}